# Controlling free electrons with optical whispering-gallery modes

Ofer Kfir[1], Hugo Lourenço-Martins[1], Gero Storeck[1], Murat Sivis[1], Tyler R. Harvey[1], Tobias J. Kippenberg[2], Armin Feist[1], and Claus Ropers[1]

[1]University of Göttingen, IV. Physical Institute, Göttingen 37077, Germany
[2]École Polytechnique Fédérale de Lausanne (EPFL), CH-1015 Lausanne, Switzerland

ofer.kfir@phys.uni-goettingen.de, claus.ropers@uni-goettingen.de

**Free-electron beams serve as uniquely versatile probes of microscopic structure and composition[1,2], and have repeatedly revolutionized atomic-scale imaging, from solid-state physics to structural biology[3–5]. Over the past decade, the manipulation and interaction of electrons with optical fields has seen significant progress, enabling novel imaging methods[6], schemes of near-field electron acceleration[7,8], and culminating in 4D microscopy techniques with both high temporal and spatial resolution[9,10]. However, weak coupling strengths of electron beams to optical excitations[11,12] are a standing issue for existing and emerging applications of optical free-electron control.**
**Here, we demonstrate phase matched near-field coupling of a free-electron beam to optical whispering gallery modes of dielectric microresonators. The cavity-enhanced interaction with these optically excited modes imprints a strong phase modulation on co-propagating electrons, which leads to electron-energy sidebands up to hundreds of photon orders and a spectral broadening of 700 eV. Mapping the near-field interaction with ultrashort electron pulses in space and time, we trace the temporal ring-down of the microresonator following a femtosecond excitation and observe the cavity's resonant spectral response. Resonantly enhancing the coupling of electrons and light via optical cavities, with efficient injection and extraction, can open up novel applications such as continuous-wave acceleration, attosecond structuring, and real-time all-optical electron detection.**

The interaction of free electrons with optical excitations forms the basis of microscopy techniques yielding fundamental insights into optical properties at the nanoscale. One of the most widespread methods, electron-energy loss spectroscopy (EELS), is extensively used to analyze resonant nanostructures[13,14], and allows for the measurement of the local photonic density of states[15–18] with nanometer precision. The acquisition of laser-driven electron-energy gain spectra (EEGS) allows for extremely accurate spatial and spectral information for nanoplasmonics[19–22], and as recently shown, for resolving the band structure mode profiles in a photonic crystal[23,24].

Inelastic scattering of electrons with optical excitations of nanostructures naturally cause changes to the free-electron state. In the case of stimulated interactions, spatial and temporal phase information of a light field is imprinted onto the electronic wavefunction passing through the optical field. A control over the wavefunction of free-electron beams was recently demonstrated in spatial[25], temporal[26] and spatiotemporal[27,28] forms. A particular example, the generation of electron pulses of attosecond-scale duration[29–31], could combine sub-optical cycle temporal resolution with the high spatial resolution of transmission electron microscopes (TEMs). The above endeavors harness the enhancement of PINEM[6] (photon-induced near field electron microscopy) from the large polarizability of plasmons. Nonetheless, a limiting factor on the strength of such couplings is typically the sub-optical cycle interaction time of swift electrons across spatially localized fields[11,13]. In analogy to other nonlinear processes, the coupling can be enhanced by coherently accumulating transition amplitude in more extended structures[32,33]. That requires phase matching, i.e., equating the electron group velocity with the optical phase velocity. Nearly relativistic electrons with energies around 100-200 keV are naturally phase-matched to traveling waves at visible frequencies in ordinary dielectrics, such as fused silica, which was previously exploited in a prism geometry for electron acceleration[34] and for stimulated Cherenkov-type interaction[35]. However, approaching a stronger coupling between photons and free electrons, ultimately leading to significant entanglement[36,37], necessitates the combination of traveling-wave phase-matching with a high density of photonic states.

Here, we address this challenge and harness whispering gallery mode (WGM) microcavities to enhance the interaction between light and free electrons. Whispering gallery modes, akin to their acoustic analogue, are traveling wave optical resonances confined in dielectric resonators by total internal reflection, achieving the highest finesse of any optical resonator to date. Leading to multiple applications, the strong field enhancements in WGM resonators allow for the radiation-pressure manipulation of mechanical modes in cavity optomechanics[38], the generation of soliton frequency combs[39] via the material's third-order nonlinearity, or near-field sensing[40] via measuring dispersive frequency shifts[41]. The WGM near field extending into free-space, which is instrumental for atomic and molecular sensing[42–44], also motivates the use of these structures for coupling to free electrons.

In the experiments presented here, we focus on two scenarios, using microspheres with diameters of 2 µm and 5 µm. The WGM in the smaller spheres allow us to reach an extremely strong and coherent modulation of the electron wavefunction, manifested in the emergence of electron-energy sidebands over 220 eV and 700 eV for chirped picosecond and unchirped femtosecond optical pulses, respectively. For the larger spheres, the smaller free-spectral range allows us to analyze the coupling of multiple WGMs to electrons, and to trace the optical cavity ringdown time. These results establish WGM as a key element in future approaches to coherently manipulate free electrons. Moreover, we believe in a high potential of WGM-enhanced electron spectroscopy, ultimately facilitating the controlled coupling to individual quantum systems.

The experiments are conducted in an ultrafast transmission electron microscope featuring electron pulses down to 200 fs in duration[45] (see illustration in Fig. 1a). These electron pulses are passed near the surface of optically illuminated silica microspheres. The WGMs are excited by optical pump pulses with a center wavelength of 800 nm. The similar velocities of the electron and the WGM's phase allow them to efficiently exchange energy through the evanescent light field that permeates vacuum (inset).

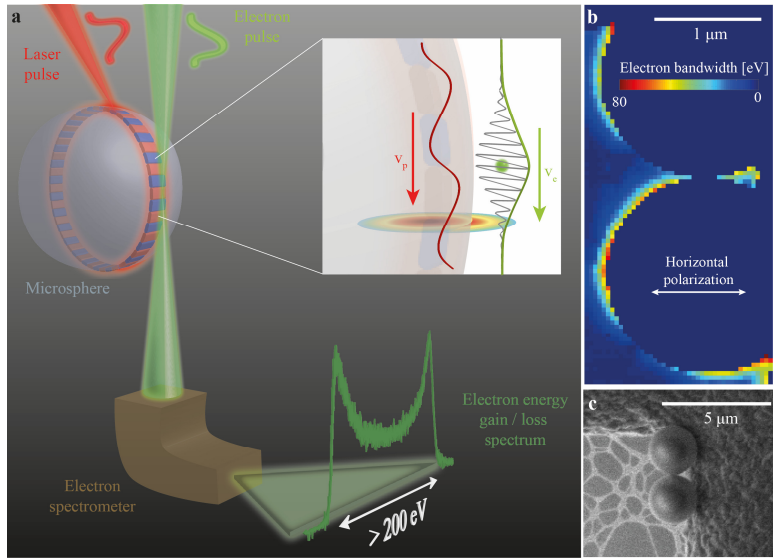

**Figure 1 | The experimental scheme**. **a** A laser-driven whispering gallery mode (WGM) circulating a microsphere drives stimulated gain and loss in a traversing electron, resulting in an exceptionally broad energy spectrum spanning hundreds of sidebands. Inset: The efficient transfer of energy and momentum is enabled by matching the velocities of the electron and the wave in the optical cavity, within the evanescent modal field. **b** A spatial map of interaction strength for two neighboring spheres, with linearly polarized excitation, quantified in terms of electron spectral bandwidth. **c** Scanning electron micrograph of the two spheres on the support structure.

Fig. 1b displays a spatial map of the electron-WGM interaction strength, with the color code representing the width of the electron energy spectrum. The spectrum substantially broadens near the surface of the sphere and to a lesser extent near metallic edges of the support structure (see methods). The interaction strength is evident along wide sections of the sphere circumference, suppressed only in regions approximately perpendicular to the incident horizontal polarization of the light. This distribution is wider than the expected width from a single WGM (which would be close to one wavelength), suggesting that a wide range of azimuthal propagation angles is excited by the illumination.

We have observed very similar halo-shaped interaction maps on multiple individual spheres and assemblies, with overall magnitudes depending on the specific geometrical arrangement and in-coupling conditions. An azimuthally selective excitation of a WGM is achieved in structures with a substantial geometrical anisotropy, such as the doublet displayed in Fig. 2a. Here, a pair of 2-µm-diameter spheres was partially merged along an axis perpendicular to the edge of the supporting copper grid (see Fig. 2a). For this structure, the interaction strength is maximized when the pump excites a WGM with transverse magnetic polarization (TM), that is, for polarization along the doublet axis. The electron-energy gain and loss via the PINEM are driven by the polarization component parallel to the electron path. The spatial dependence of the resulting electron bandwidth in Fig. 2b shows very good agreement with the expected PINEM interaction for a single TM-polarized WGM, as simulated in Fig. 2c. Specifically, both the azimuthal extent of about one wavelength and the exponential radial decay with a characteristic scale of $L_{decay}^{exp} = L_{decay}^{calc} = 0.1$ µm are reproduced. The experimental evaluation of the decay length is based on a line scan, for which the spectral bandwidth was estimated for each point as a function of its distance from the surface (orange circles in Fig. 2d). The individual curves in Fig. 2d are spectra recorded for a set of fixed distances, which visualizes the broadening of the spectrum near the sphere. Close to the sphere's surface, the spectral bandwidth extends to a remarkable 220 eV, while keeping the double-lobed shape characteristic to a PINEM spectrum for a single, homogeneous interaction strength, corresponding to a high-purity electron-energy comb[26]. In this experiment, by temporally stretching the optical pulse to 3.5 ps FWHM (full-width-at-half-maximum), the driving field amplitude is essentially constant during the interaction with the electron pulse, and thus the phase modulation imprinted on the electron wave function is uniform and deterministic[26]. Pumping the system with shorter pulses of 400 fs duration enables an increase of the optical field strength, albeit at the cost of some variation of the coupling strength across the electron pulse duration. The top-left inset shows an electron spectrum with a significantly wider bandwidth of 700 eV, far beyond what is typically observed in PINEM[26,29,46].

To quantify the temporal uniformity of the interaction, we fit the experimental data to the expected PINEM spectrum[26,32] with normally distributed values of the interaction parameter, with a mean value of $g$ and a standard deviation $\Delta g$. This accounts for both spatial and temporal averaging of the PINEM signal. The relative uncertainty of $g$, that is, $\frac{\Delta g}{g}$, can be regarded as reducing the purity of the final phase-modulated electron state. In the limit of $|g| \gg 1$, the electron-energy width (or bandwidth) scales as $4|g|$ times the optical photon energy (1.55 eV for $\lambda = 800$ nm light). An ideal PINEM spectrum, with $\Delta g/g = 0$ is composed of discrete orders separated by the photon energy, with relative probabilities given by the Bessel function of the first kind, $P_k = |J_k(2|g|)|^2$, where $k$ is the gain/loss sideband order. We evaluated the relative uncertainty of the interaction strength for the 220-eV-wide and the 700-eV-wide spectra as $\frac{\Delta g}{g} = 0.09$, and $\frac{\Delta g}{g} = 0.33$, respectively. The temporal uniformity of the driving field across the electron pulse is a crucial parameter for the quality of the final state prepared by the interaction. For example, $g$ defines the propagation length scale at which electron phase modulation evolve to probability modulations.

Thus, $\frac{\Delta g}{g} \ll 1$ is required to maintain a well-defined evolution of the wavefunction, and to maintain the coherence of the phase modulations imprinted onto the electron.

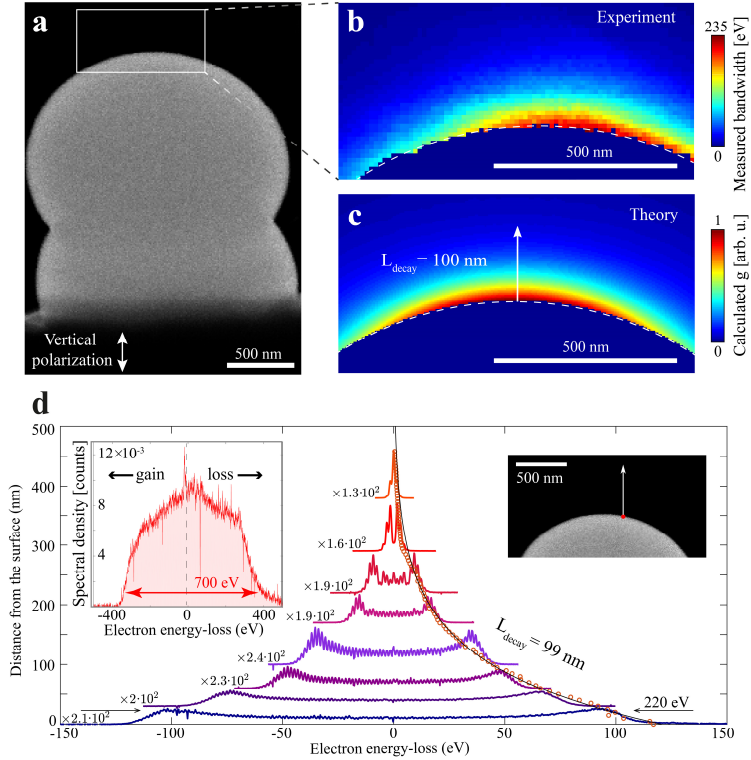

**Figure 2 | Electron spectral broadening induced by whispering-gallery modes**. **a** An annular dark-field (ADF) image of a microsphere doublet comprising a pair of 2 μm spheres, optically pumped from the top (into the page plane). **b** Measured electron bandwidth as a function of position. **c** Simulated electron bandwidth from the coherent interaction with a single WGM. **d** Electron spectra measured at a varying distance from the sphere's surface, reaching a bandwidth of 220 eV with a state purity of 91% (bottom curve, see text). The use of a shorter 400-fs-long pulse extends the electron bandwidth to a value of 700 eV (left inset), albeit at a reduced purity (67%). The measured interaction strength (red circles) exhibits a very good agreement to an exponential (solid line) with a 100 nm decay length.

In order to characterize the resonance properties in the interaction with the WGMs, we utilize larger spheres with a diameter of 5 μm (see Fig. 3a), for which the optical bandwidth is larger than the free-spectral range, and thus covers several discrete cavity modes. We first study the temporal decay of the cavity field by measuring the electron spectrum as a function of time delay between the electron pulses and 50-fs optical pump pulses (see Fig. 3b). The electron spectrum rapidly broadens around the temporal overlap (t=0) and narrows down to its original width with an exponential decay time of about 260 fs (cf. Fig. 3b, right panel), as expected for a Lorentzian resonance with a quality factor of $Q = 97$. The fact that this value is substantially below the calculated radiative quality factor ($Q = 1230$) is likely caused by scattering from structural imperfections and the support, suggesting the possibility for improvements in optimized geometries. The significant uncertainty of the coupling, $\Delta g \approx g$, points to the presence of more than one whispering gallery mode and corresponding modal beatings which are faster than the electron pulse duration.

For a spectral mode analysis, we pump the microsphere with a strongly chirped pulse, mapping optical frequencies to arrival times. The delay-dependent electron spectrogram (Fig. 3c) now shows three broadened regions, corresponding to distinct WGM indices. By comparison with the computed time-dependent cavity response for the chirped pulse (dash line, arbitrary units), we find that the strong features represent two transverse magnetic (TM) modes with vacuum wavelength $\lambda_{\ell=22} = 804\ nm$ and $\lambda_{\ell=23} = 773\ nm$. The weaker is attributed to a transverse electric (TE) mode at $\lambda_{\ell=23} = 790\ nm$, following from a lower field along the electron trajectory (see color-coded mode profiles of the relevant out-of-plane field component, $E_\phi$). Aside from the agreement on the general features, the calculation also predicts temporal oscillations in the response, as the frequency sweep is not completely adiabatic with respect to the cavity decay rate. Similar oscillations are indeed discernible in some areas of the experimental spectrogram between t=0 and t=1 ps.

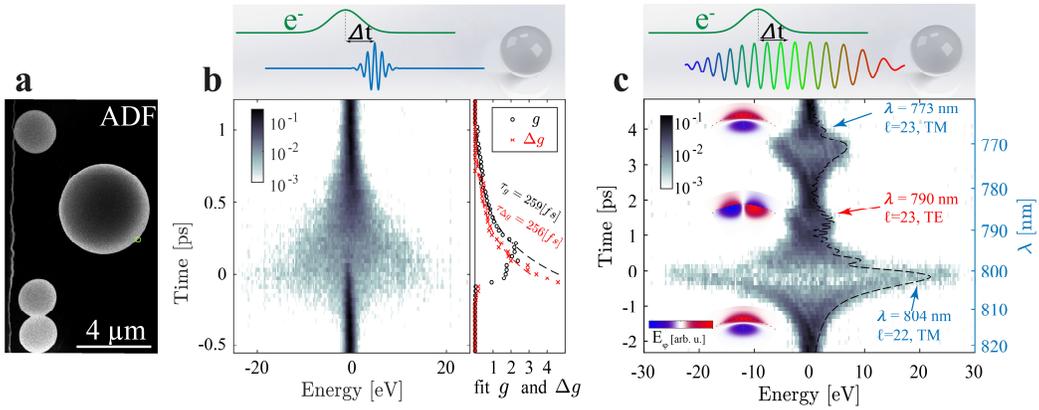

**Figure 3 | Spectral and temporal properties of free-electron interaction with whispering-gallery modes. a** Annular dark-field micrograph of free-standing spheres with 2 μm and 5 μm diameters. The yellow circle marks the measurement position. **b** (left) Colormap of the electron spectrum (log-scale) following a loading of WGMs by a 50-fs-long optical pulse at time = 0. At later times, the electron is driven by light stored in the cavity, with a storage time of 260 fs, corresponding to a quality factor Q = 97. The lifetime is evaluated from fitting $g$ and $\Delta g$ with an exponential. **c** The WGM selectivity probed by the electron's spectral response. Using a frequency sweep (chirped pulse), three modes are identified as an increased electron bandwidth at delay times that match the sweep to their resonance frequency. The simulated cavity response (dashed black) agrees with the experiment, which allows us to identify the mode polarizations and indices (see marks) for a sphere diameter of D = 4.765 μm. The colormap presents the PINEM-relevant azimuthal component, $E_\phi$, of the identified modes.

In conclusion, our work demonstrates an efficient coupling of free-electron beams to whispering-gallery mode microresonators, leading to an electron spectral broadening up to 700 eV. While the current study focuses on low-quality factors (Q) and free-space excitation, it can readily be extended to photonic-chip based microresonators (e.g. based on $Si_3N_4$) operating with Q factors in excess of $10^7$ [47] and using phase-matched, high-ideality and fiber pigtailed nanophotonic waveguides for excitation. Importantly, such an approach not only allows for the excitation of WGM, but also close-to-unity collection efficiency of the light that has interacted with electrons. Moreover, attosecond optical modulation of the phase and density of continuous electron beams in standard electron microscopes appears in reach. Such high-current electron beams dressed by light could transfer optical polarizations down to nanometer-sized focal spots, possibly acquiring high resolution spectroscopic information from resonator-coupled atoms[42,43,48], molecules[40,44] and nanoparticles[49]. In a similar vein, continuous wave probing of the resonators' phase response could enable all-optical real-time detection of electrons. The ability of microresonators to generate optical dissipative solitons may allow for the coupling of electrons to tightly localized fields only few optical cycles long[39], relevant for time-gated interaction. More generally, cavity-enhanced and phase-matched near-field-electron interactions promise a merging of electron microscope and photonic chip-based microresonators, with far-reaching consequences in local quantum control and sensing.

## Methods

**Experimental details**
The experimental system is a transmission electron microscope (JEOL JEM 2100F) modified to allow for a pulsed photo-electron beam, with pulses as short as 200 fs. The technical details are elaborated in Ref. [45]. The pump laser is an amplified Ti:Sapphire system (Coherent RegA), providing pulses centered at a wavelength of 800 nm, full-width at half maximum (FWHM) bandwidth of 34 nm, a repetition rate of 600 kHz, and an average power of 150 mW entering the TEM. Additional glass bars (19-cm-long SF6, 10-cm-long N-BK7, or none) can be placed in the beam path to add temporal dispersion. The optical beam is nearly co-propagating with the electron beam (6° off axis) and focuses down to a characteristic mode size of 10 μm FWHM. The timing between the electron pulse and the laser is controlled by a delay stage.
The experiment (see illustration in Fig. 1a) investigates the electron interaction with WGMs circling in silica microspheres with diameters of 2 μm and 5 μm (SSD5003 and SSD5000, respectively, Bangs Laboratories Inc.). For sample preparation, the spheres were immersed in ethanol and randomly distributed by drop-casting on a Lacey-carbon support (Ultrathin Carbon / Lacey Support on 400 mesh, Ted Pella, Inc.). The experiments utilize Scanning TEM (STEM) for a systematic acquisition of the electron spectrum and spectral bandwidth (Gatan "Enfinium" spectrometer) and an annular dark field (ADF) detector.

**Simulation details of the PINEM map, Fig. 2c**
The calculation uses electrons accelerated to 200 keV, locally interacting with a WGM in a 2 μm sphere, having a vacuum wavelength of $\lambda = 806\ nm$, which was found to be the mode closest to the laser wavelength using the WGMode toolbox[50]. Furthermore, the WGMode toolbox was used to calculate the $E_\phi$ component of the electric field on the equatorial plane, from which the PINEM-relevant field in space and time was calculated by adding the temporal and azimuthal phase $e^{i(\omega t - \ell \phi)}$. The contribution to the PINEM signal at each coordinate $(x, y)$, was integrated over the electron trajectory, $t(z)$, as $\int_{-\infty}^{\infty} E_\phi(x, y, z, t_{(z)}) dz$. The other field components do not contribute.

**Calculation of the expected spectral response of the WGM in Fig. 3c**

The only fitting parameter for the expected WGM-driven electron spectrum (dashed line on Fig. 3c) is the sphere's diameter, for which the best fit was found at D=4.765. This value is only 2% smaller than the measured diameter. To evaluate the resonator response, we used the tabulated index of refraction for fused silica, n=1.4533[51], and the experimentally measured lifetime 260 fs (see Fig. 3b). The wavelength of the resonances was calculated by the WGMode package[50], and the relative strength of each of the resonances was based on the maximal field components, $E_\phi$, of the different modes. The TM modes are maximal at the center, and the TE mode at a slightly shifted position compared with the plane of circumference (see red and blue colormaps on Fig. 3c). The arrival time of the central pumping wavelength ($\lambda = 800\ nm$) is calibrated based on PINEM experiments on metallic surfaces and is determined as time zero.

**Data Availability**

The data that support the findings of this study are available from the corresponding author upon reasonable request.

**Acknowledgements**
O.K. gratefully acknowledges funding from the European Union's Horizon 2020 research and innovation programme under the Marie Skłodowska-Curie grant agreement No.752533. This work was funded by the Deutsche Forschungsgemeinschaft (DFG) in the Collaborative Research Center "Atomic Scale Control of Energy Conversion" (DFG-SFB 1073, project A05) and in the Priority Program "Quantum Dynamics in Tailored Intense Fields" (DFG-SPP 1840).